\documentclass[11pt]{article}
\usepackage{mathrsfs}
\usepackage{graphicx}  
\usepackage{float}
\usepackage{amssymb}
\usepackage{amsmath}
\usepackage{cases}
\usepackage{amsfonts}
\usepackage{mathrsfs}

\newtheorem{theorem}{Theorem}[section]

\newtheorem{corollary}[theorem]{Corollary}
\newtheorem{lemma}[theorem]{Lemma}
\newtheorem{alg}[theorem]{Algorithm}

\newtheorem{defi}[theorem]{Definition}
\newtheorem{example}[theorem]{Example}

\def\qed{\hfil {\vrule height5pt width2pt depth2pt}}
\def\bref#1{(\ref{#1})}

\def\C{\mathcal{C}}

\def\Cont{{\rm{Cont}}}
\def\coeff{\hbox{\rm{coeff}}}
\def\Prim{\hbox{\rm{Prim}}}

\def\bref#1{(\ref{#1})}

\def\Q{{\mathbb Q}}

\def\M{{\mathcal M}}

\def\P{{\mathscr{P}}}

\def\bref#1{(\ref{#1})}

\def\Z{{\mathbb Z}}
\def\C{{\mathbb C}}
\def\M{{\mathbb M}}

\def\+{ \oplus}
\def\-{\ominus}
\def\*{\otimes}

\def\C{{\mathbb{C}}}

\title{Multiplicity Preserving Triangular Set Decomposition of Two Polynomials}

\author{Jin-San Cheng, \hspace{1cm}
 Xiao-Shan Gao\\
Key Lab of Mathematics Mechanization\\ Institute of Systems
Science, AMSS, Chinese Academy of Sciences\\
jcheng@amss.ac.cn, xgao@mmrc.iss.ac.cn}

\date{}

\begin{document}
\maketitle
\begin{abstract}
In this paper, a multiplicity preserving triangular set
decomposition algorithm is proposed for a system of two polynomials.
The algorithm decomposes the variety defined by the polynomial
system into unmixed components represented by triangular sets, which
may have negative multiplicities. In the bivariate case, we give a
complete algorithm to decompose the system into multiplicity
preserving triangular sets with positive multiplicities. We also
analyze the complexity of the algorithm in the bivariate case. We
implement our algorithm and show the effectiveness of the method
with extensive experiments.

\end{abstract}

\section{Introduction}

Decomposing a polynomial system into triangular sets is a classical
method to solve polynomial systems. The method was first introduced
by Ritt \cite{ritt} and revised by Wu in his work of elementary
geometry theorem proving \cite{wu1,wu2}. There exist many work about
this topic
\cite{alm,chen,chou,dahan,gao,lazard,lazard1,lixin,mmm,kal,kalaaecc,kal2,wang,yangzhang}.
The main tool to decompose a polynomial system is pseudo-division.
In most existing triangular decomposition methods based on the
pseudo-division algorithm,  one need to deal with the initial of
certain polynomial(s), say $h$, which will bring extraneous zeros.
Usually, one decomposes here the system into two systems corresponding to the cases $h=0$
and $h\neq 0$. Doing so, the number of the systems increases
quickly. Moreover, this leads to some repeated computations which
can be avoided.

Another reason why we consider the topic is that the multiplicity of a component or
a zero of a polynomial system is an important information which
helps us to obtain a further understanding of the structure of the
variety defined by the polynomial system.

Most triangular set decomposition algorithms do not preserve the multiplicities
of the zeros or the components.
One approach to remedy this drawback is to decompose the polynomial
system into triangular sets first and then recover the
multiplicities.
Li proposed a method to compute the multiplicities
of zeros of a zero-dimensional polynomial system after obtaining a
triangular decomposition of the system \cite{li}.
Recently, Li, Xia, and Zhang proved that the characteristic sets in
Wu's sense for zero-dimensional polynomial system is actually
multiplicity preserving with a minor modification \cite{xia}. They also gave a multiplicity
preserving decomposition, but some of the components are not in
triangular form.

In this paper, we use the concept of multiplicative variety, that
is, the components and their multiplicities in the original
polynomial system. We consider not only the components themselves
but the multiplicities of these components.
During the decomposition, the initials bring some extraneous
multiplicative varieties in each pseudo-division step. We record
them during the computation and remove them later, which helps us to
recover the multiplicative varieties of the original system. We also
avoid some repeated computation during the decomposition. Currently,
the theory is complete for polynomial system with two polynomials.
In particular, we provide a method to compute the multiplicative-zeros
of a zero-dimensional bivariate system with two polynomials. We also
analyze the complexity of the algorithm under some conditions.

Kalkbrener's method for zero-dimensional bivariate polynomial system is
similar to our method \cite{kalaaecc}. But his method is not
multiplicity preserving. And our method is in a different sense: we
remove the extraneous zeros from the system.

The paper is organized as below.
In the next section, we provide
some properties of primitive polynomial remainder sequences. In
Section 3, we provide the theories to decompose a polynomial system
with two polynomials into triangular sets which preserve the
multiplicities of the components of the original system. We provide
a multiplicity preserving algorithm to decompose a zero-dimensional
bivariate polynomial system into triangular sets in Section 4. The
complexity of the algorithm under some conditions are analyzed.
Algorithms and examples are used to illustrate the effectiveness
and efficiency of our method. We also compare our method with other
related methods. We draw a conclusion in the last section.

\section{Primitive Polynomial Remainder Sequence}
In this section, we introduce some basic properties for primitive
polynomial remainder sequences. In fact, there are many references
for this topic, in particular \cite{brown,kal,kalaaecc}. We modify
the procedure for our own purpose.

Let $K$ be a computable field with characteristic zero, such as the
field of rational numbers and $K[y_1,\ldots,y_n]$ the polynomial
ring in the indeterminates $y_1,\ldots,y_n$.

Let $p\in K[x_1,\ldots,x_n,x]$. We define
\begin{eqnarray*}
\Cont(p,x)&=& \gcd(\coeff(p,x^i),i=0,1,\ldots,\deg(p,x)),\\
\Prim(p,x)&=& p/\Cont(p,x),
\end{eqnarray*}
where $\coeff(p,x^i)$ means the coefficient of $x^i$ in $p$ and $\deg(p,x)$ means the degree of $p$ in $x$. $p$ is called {\bf primitive } w.r.t. $x$ if $\Cont(p,x)=1$.

The pseudo-division can be extended to the following form.

\begin{lemma}\label{lm-q}
Let $f,g\in K[x_1,\ldots,x_n,x]$, $\deg(f,x)=d_1$, $\deg(g,x)=d_2$,
$d_1\ge d_2$, and $\gcd(f,g)=1$. There exist $q,
r\in K[x_1,\ldots,x_n,x]$ such that
\begin{equation}\label{eq-3}
l^{\delta+1} f+q\, g=r,
\end{equation}
 where $l$ is the leading coefficient of $g$ in $x$, $\delta=d_1-d_2$, $\deg(g,x)>\deg(r,x)$. Furthermore, $q$ has the form:
\begin{equation}\label{eq-2}
q=l\, t\,x+s,
\end{equation}
where $t\in K[x_1,\ldots,x_n,x], s\in K[x_1,\ldots,x_n]$. Moreover, if $r_1=\Cont(f,x),r_2=\Cont(g,x)$, then
\begin{equation}\label{eq-q}
r_1|q,\,r_2^{d_1-d_2}|q, r_1|r, r_2^{d_1-d_2+1}|r.
\end{equation}
\end{lemma}
{\bf Proof.} Write $f, g$ as univariate polynomials in $x$,
\begin{eqnarray*}
f&=&a_1\,x^{d_1}+a_2\,x^{d_1-1}+\cdots+a_{d_1+1},\\
g&=&b_1\,x^{d_2}+b_2\,x^{d_2-1}+\cdots+b_{d_2+1}.
\end{eqnarray*}
To eliminate the terms of $f$ with degree $d_1$ in $x$, we have
$$T_0(x)=b_1\,f+ q_0\, g=h_0\,x^{d_1-1}+{\hbox{ lower powers in x}},$$
where $q_0=- a_1\,x^{d_1-d_2}, h_0=b_1\,a_2 - a_1 b_2$. It is clear that $r_1|q_0$, since $r_1|a_1$, $r_2^0(=1)|q_0$ and $r_1|T_0, r_2|T_0$. So the lemma holds when $\delta=0$. Note that $r=T_0$ when $\delta=0$.
Now, we need to eliminate $h_0*x^{d_1-1}$ from $T_0(x)$. If $h_0\neq 0$,
\begin{eqnarray*}
T_1(x)&=&b_1\,T_0(x) - (b_1\,a_2 - a_1 b_2)\,x^{d_1-d_2-1} g\\
&=& b_1^2\,f+(b_1\,q_0-(b_1\,a_2 - a_1 b_2)\,x^{d_1-d_2-1}) g\\
&=& b_1^2\,f+q_1\, g\\
&=&h_1\,x^{d_1-2}+{\hbox{ lower powers in }}x,
\end{eqnarray*}
where $h_1\in K[x_1,\ldots,x_n]$. Each term of $q_1$ contains a factor of the form $a_i\,b_j$. So $r_1|q_1$, $r_2|q_1$ and $r_1|T_1, r_2^2|T_1$. And $q_1=-b_1\,a_1\,x^{d_1-d_2}-(b_1\,a_2 - a_1 b_2)\,x^{d_1-d_2-1}$. If $h_0=0$, the results is still true. So the lemma holds when $\delta=1$.
Assuming that the lemma holds for the cases $\delta\le i$, then we have
\begin{eqnarray*}
T_j(x) &=& b_1^{j+1}\,f+q_j\, g = h_j\,x^{d_1-j-1}+{\hbox{ lower powers in }}x,\\
q_j &=& b_1 q_{j-1}-h_{j-1}x^{d_1-d_2-j},\\
&& r_1|q_j,r_2^j|q_j,r_1|T_j,r_2^{j+1}|T_j, j\le i.
\end{eqnarray*}
Note that $\deg(q_{j-1},x) > \deg(q_j,x)$ and the lowest power of $q_{j-1}$ in $x$ is larger than $d_1-d_2-j$ and $r_1|h_j,r_2^{j+1}|h_j$. Then
$$T_{i+1}(x) = b_1^{i+2}\,f+q_{i+1}\, g = h_{i+1}\,x^{d_1-i-2}+{\hbox{ lower powers in x}},$$
We can similarly derive $q_{i+1}=b_1 q_i-h_i\, x^{d_1-d_2-i-1}$. When $\delta=i+1$, we have $q_{i+1}=b_1 q_i-h_i$. So
$q_{i+1}$ has form \bref{eq-2} since the lowest power of $q_i$ is larger than $0$. And $r_1|q_{i+1}$ since $r_1|q_i, r_1|h_i$. So $r_1|T_{i+1}(x)$. $r_2^{i+1}|q_{i+1}$ since $r_2|b_1,r_2^i|q_i$ and $r_2^{i+1}|h_i$. So $r_2^{i+2}|T_{i+1}(x)$. So the lemma holds for $\delta=i+1$.
The lemma is proved.
\qed

\begin{corollary}\label{lem-eu}
Let $f,g\in K[x_1,\ldots,x_n,x]$ be primitive, $d_1=\deg(f,x)\ge
d_2= \deg(g,x)$, and $\gcd(f,g)=1$. Regard $f, g$ as univariate
polynomials in $x$. Then there exist an $m\in K[x_1,\ldots,x_n]$
such that
$$m\,f=q\,g+r,$$
and $\gcd(m,q)=\gcd(m,r)=1$. Furthermore,
$$(m\,f,g)=(g,r),$$
where $(P)$ represents the ideal generated by $P$.
\end{corollary}
{\bf Proof.}
The corollary is obvious. \qed

\begin{corollary} 
$\Cont(g,x)=1$ if $\gcd(l, s)=1$ and $d_1 > d_2$, where $l$ and $s$ are from \bref{eq-3} and \bref{eq-2}.
\end{corollary}
{\bf Proof.} Regard $f, g$ as univariate polynomials in $x$, and $q$ a polynomial in $x$ and $a_i,b_j$, where $i=1,\ldots,d_1+1, j=1,\ldots,d_2+1$. Let $r_2|g$ and $r_2\in K[x_1,\ldots,x_n]$. From Lemma \ref{lm-q}, $r_2|q$ if $d_1 > d_2$. So $r_2|s$. Since $r_2|l$, $r_2|\gcd(l,s)$. We have $r_2=1$ if $\gcd(l,s)=1$. The lemma is proved.
\qed

The above result is a necessary condition to check whether $g$ has factors in $K[x_1,\ldots,x_n]$.

\begin{lemma}\label{lm-eud}
Let $f_1,f_2\in K[x_1,\ldots,x_n,x]$, $d_1=\deg(f_1,x)$ $\ge d_2=\deg(f_2,x)$. Assume that $\Cont(f_i,x)=1, i=1,2$. Applying the extended Euclidean algorithm for $f_1,f_2$ w.r.t. the variable $x$, we obtain a polynomial sequence $\{f_1,f_2,\ldots$, $f_{k+2}\}$ such that
\begin{eqnarray} \label{eq-n1}
m_i f_i+q_i f_{i+1} &=& m_{i-1} p_i f_{i+2}, i=1,\ldots,k, \label{eq-n1}
\end{eqnarray}
where $m_0=1, p_k=1, m_i,p_i,f_{k+2}\in K[x_1,\ldots,x_n], q_i\in K[x_1,\ldots,x_n,x], i=1,\ldots,k$, and $\Cont(f_i,x)=1 (1\le i\le k+1)$, $\gcd(m_i,p_i)=1$.
\end{lemma}
{\bf Proof.} We prove the lemma by induction on $i$. When $i=1$, from Lemma \ref{lm-q}, there exist $q\in K[x_1,\ldots,x_n,x]$, $r\in K[x_1,\ldots,x_n,x]$ such that $l_2^{\delta+1} f_1+q f_2=r$, where $l_i$ is the leading coefficient of $f_i$ in $x$, $\delta=d_1-d_2$. Let $t=\gcd(l_2^{\delta+1},q)$, $m_1=\frac{l_2^{\delta+1}}{t}, q_1=\frac{q}{t}$. Let $p_1=\frac{\Cont(r,x)}{t}$ and $f_3=\Prim(r,x)$. It is clear that $\gcd(m_1,p_1)=1$. Assume that for $1\le j< i$, \bref{eq-n1} holds. Denote $d_i=\deg(f_i)$. For $j=i$, we have $l_{i+1}^{\theta+1} f_i+q_t f_{i+1}=r_{i+2}$, where $\theta=d_i-d_{i+1}$. If $m_{i-1}$ is a factor of $r_{i+2}$,
set $p_i'$ as the product of all the factors of $\frac{r_{i+2}}{m_{i-1}}$ in $K[x_1,\ldots,x_n]$. Let $h=\gcd(l_{i+1}^{\theta+1},p_i')$. Then $m_i=\frac{l_{i+1}^{\theta+1}}{h},q_i=\frac{q_t}{h}, p_i=\frac{p_i'}{h}$, $\gcd(m_i,p_i)=1$. If $m_{i-1}$ is not a factor of $r_{i+2}$, we can multiply $g=\frac{m_{i-1}}{\gcd(m_{i-1}, r_{i+2})}$ to the two sides of the equation. Then doing the same operation as before, we can derive $m_i f_i+q_i f_{i+1} = m_{i-1} p_i f_{i+2}$ which satisfies all the conditions. This proves the lemma.
\qed

\noindent{\bf Remark: } In most cases, we have $\gcd(m_i,q_i)=1$ and $p_i=1$ which helps us to design efficient algorithms.

\begin{corollary}\label{cor-8}
Let $f_1,f_2\in K[x_1,\ldots,x_n,x], \gcd(f_1,f_2)=1$, and $\Cont(f_i,x)=1, i=1,2$.  From the extended Euclidean algorithm, we can obtain
\begin{eqnarray}
m_i f_i+q_i f_{i+1}&=& g_i f_{i+2},\,\, i=1,\ldots,k, \label{eq-8}\\
(m_i\,f_i,f_{i+1})&=&(f_{i+1},g_i\,f_{i+2}),\label{eq-9}
\end{eqnarray}
where $m_i,g_i\in K[x_1,\ldots,x_n]$,  $\gcd(m_i,g_i)=1$,  $g_k=1$,  $\gcd(m_k,g_k f_{k+2})=1$, and $f_{i+2}(1\le i\le k-1)$ are primitive.
\end{corollary}
{\bf Proof.} From Lemma \ref{lm-eud}, we have \bref{eq-n1}. Note that $\gcd(m_i,p_i)=1$. Let $h=\gcd(m_i,m_{i-1})$, denote $m_i=\frac{m_i}{h}, q_i=\frac{q_i}{h}, g_i=\frac{m_{i-1}p_i}{h}$. Then we have $\gcd(m_i,g_i)=1$. Since $f_{k+2}\in K[x_1,\ldots,x_n]$, we can set $g_k=1$. We can delete $\gcd(m_k,g_k f_{k+2})$ if it exists. \bref{eq-9} is obvious. So the corollary holds.
\qed

 The following corollary is clear and useful.
\begin{corollary}\label{cor-4}
 We can rewrite \bref{eq-8} and \bref{eq-9} as below.
\begin{eqnarray}
m_i f_i+q_i f_{i+1}&=& \frac{m_{i-1}}{w_i}p_i f_{i+2}, i=1,\ldots,k,\label{eq-18}\\
(m_i\,f_i,f_{i+1})&=&(f_{i+1},\frac{m_{i-1}}{w_i}p_i\,f_{i+2}),\label{eq-19}
\end{eqnarray}
where $w_i$ is a factor of $m_{i-1}$, $g_i=\frac{m_{i-1}}{w_i}p_i$, and $p_k=1$.
\end{corollary}

\section{Triangular Decomposition of Two Polynomials}
In this section, we will give the method to decompose a system of
two polynomials into triangular sets. We need the concept of
multiplicity variety.
\begin{defi}(\cite{hp} pp. 129-130) \label{def-m}
Let $V_d$ be an unmixed variety 
of dimension $d$ in a projective space $S_n$ of dimension $n$ over $K$. And
$$V_d=\sum_{i=1}^h V_d^{(i)},$$
where $V_d^{(i)}$ is an irreducible variety of dimension $d$ and order $g_i$. Let $F_i(u_0,\ldots,u_d)$ be the Chow form (see \cite{hp} pp.32) of $V_d^{(i)}$; which is irreducible over $K$
and of degree $g_i$ in $u_j=(u_{j0},\ldots,u_{jn})$, for $j=0,\ldots,d$. The form
\begin{equation}\label{chowform} F(u_0,\ldots,u_d)=\Pi_{i=1}^h[F_i(u_0,\ldots,u_d)]^{a_i},\end{equation}
where $a_1,\ldots,a_h$ are positive integers, satisfies the conditions for a Chow form of an algebraic variety which, regarded as a set of points, coincides with $V_d$. We consider a new entity, consisting of the variety $V_d$ associated with the form $F(u_0,\ldots,u_d)$, for a given choice of the exponents $a_1,\ldots,a_h$, denoted as $\M_d$. We write
\begin{equation}\label{mv}
\M_d=\sum_{i=1}^h a_i\M_d^{(i)},
\end{equation}
where $\M_d^{(i)}$ corresponds to $V_d^{(i)}$.
We call $\M_d$ a {\bf multiplicative variety} and $a_i$ the {\bf multiplicity} of $\M_d^{(i)}$. Especially, we call $\M_d$ a {\bf multiplicative-zero} set when $d=0$.
\end{defi}
\noindent{\bf Remark:}
Since an affine variety can be easily transformed into a projective variety, we will consider directly affine multiplicative varieties in $K^n$ in this paper. And we assume that the system has no solutions at $\infty$.

\begin{theorem}(\cite{hp} pp160) \label{thm-hp}
$\M_a, \M_b', \M_b''$ are unmixed multiplicative varieties with dimension $a, b$ respectively. The intersection of $\M_a$ and $\M_b'+\M_b''$ are with dimension $a+b-n$. So are the intersections of $\M_a$ and $\M_b'$, $\M_a$ and $\M_b''$. Then we have
$$\M_a \cdot (\M_b'+\M_b'')=\M_a \cdot \M_b'+\M_a \cdot \M_b'',$$
where $\cdot$ represents the intersection of two multiplicative varieties, which preserves the multiplicities of each intersection component (for more details see \cite{hp}, pp158-160).
\end{theorem}

\begin{defi}(\cite{cox} pp.139)
Let $I$ be a zero dimensional ideal in $K[x_1,\ldots,x_n]$ such that the variety $V(I)$ defined by $I$ consists of finitely many points in $\overline{K}^n$, where $\overline{K}$ is algebraic closure of $K$, and assume $p=(a_1,\ldots,a_n)\in V(I)$. Then the {\bf multiplicity} of $p$ as a zero of $I$, denoted by $m(p)$, is the dimension of the ring obtained by localizing $\overline{K}[x_1,\ldots,x_n]$ at the maximal ideal $M=I(p)=(x_1-a_1,\ldots,x_n-a_n)$ corresponding to $p$, that is:
$$m(p)={\rm dim}_{\overline{K}} \overline{K}[x_1,\ldots,x_n]_{M}/I \overline{K}[x_1,\ldots,x_n]_{M}.$$
\end{defi}

\begin{lemma}(\cite{cox} pp. 144 )
Let $\{f_1,\ldots,f_n\}\subset K[x_1,\ldots,x_n]$ be
zero-dimensional, and have total degrees at most $d_1,\ldots,d_n$
and no solutions at $\infty$. If $f_0=u_0+u_1\,x_1+\ldots+u_n\,x_n$,
where $u_0,\ldots,u_n$ are independent variables, then there is a
nonzero constant $C$ such that the Chow form of $(f_1,\ldots,f_n)$
is
\begin{eqnarray*}
&&{\rm Res}_{1,d_1,\ldots,d_n}(f_0,\ldots,f_n)\\
&=&C \Pi_{p\in V_{\overline{K}}(f_1,\ldots,f_n)}
(u_0+u_1\,\xi_1+\ldots+u_n\,\xi_n)^{m(p)},
\end{eqnarray*}
where $p=(\xi_1,\ldots,\xi_n)\in V_{\overline{K}}(f_1,\ldots,f_n)$
and $m(p)$ is the multiplicity of $p$ in $(f_1,\ldots,f_n)$.
\end{lemma}

The lemma illustrates the relationship between Chow form and the
multiplicity of a point of a zero-dimensional polynomial system. The
lemma tells us that $m(p)$ is the multiplicity of the corresponding
irreducible zero-dimensional component of the zero-dimensional
polynomial system $\{f_1,\ldots,f_n\}$  when $d=0$ in Definition
\ref{def-m}.

From Theorem \ref{thm-hp}, we have the corollary below.
\begin{corollary} \label{for-2} Using the notations as Corollary \ref{lem-eu}, we have
\begin{equation*}
\M(m\,f, g)=\M(m,g)+\M(f,g).
\end{equation*}
\end{corollary}

The following lemma is important to our algorithm.
\begin{lemma}\label{thm-decom}Let $f,g,r,m$ be as Corollary \ref{lem-eu}.
We have
\begin{equation}\label{for-decom}
\M(f,g)=\M(g,r)-\M(m,g).
\end{equation}
\end{lemma}
{\bf Proof.} By Corollary \ref{lem-eu}, $(m\,f,g)=(g,r)$, so we have
\begin{equation}\label{for-1}
\M(m\,f, g)=\M(g,r).
\end{equation}

We can find that $m\,f,g$ define a multiplicative variety with
dimension $n-2$ since $\gcd(m\,f,g)=1$.
Note that $f,g$ both are primitive and $m\in K[x_1,\ldots,x_n]$.
And $f, g$ and $m,g$ both
are $n-2$ dimensional.

We are going to prove that
\begin{equation}\label{for-3}
\M(f,g)=\M(m\,f, g)-\M(m,g).
\end{equation}

Assume that the Chow form $T_0$ of $\M(m\,f,g)$ is as
\bref{chowform} and the order of $\M(m\,f,g)$ is $G_0=\sum_{i=1}^h
a_i\,g_i$. From \bref{for-2}, we can assume that the Chow form of
$\M(m,g), \M(f,g)$ are $T_1=\Pi_{i=1}^k[F_i(u_0,\ldots,u_d)]^{a_i}$
and $T_2=\Pi_{i=k+1}^h[F_i(u_0,\ldots,u_d)]^{a_i}$, respectively,
where $1<k<h$ and $d=n-2$. And the orders are $G_1=\sum_{i=1}^k
a_i\,g_i$ and $G_2=\sum_{i=k+1}^h a_i\,g_i$, respectively. We define
the Chow form $T_2'$ of $\M(m\,f, g)-\M(m,g)$ as $T_0/T_1$ and the
order as $G_2'=G_0-G_1$. Since the Chow form of the algebraic
variety is not equal to zero, the definition is well defined. Here
$(m,g)$ defines some components of $(m\,f,g)$, including the
multiplicities of the components. Thus $\M(m\,f, g)-\M(m,g)$ still
has positive exponent for each simplified component. And we have
\begin{eqnarray*}
T_2'&=&\Pi_{i=1}^h[F_i(u_0,\ldots,u_d)]^{a_i}/(\Pi_{i=1}^k[F_i(u_0,\ldots,u_d)]^{a_i})
=\Pi_{i=k+1}^h[F_i(u_0,\ldots,u_d)]^{a_i}=T_2,\\
G_2'&=&\sum_{i=1}^h a_i\,g_i-\sum_{i=1}^k a_i\,g_i=\sum_{i=k+1}^h a_i\,g_i=G_2.
\end{eqnarray*}
So \bref{for-3} holds.  Combining \bref{for-1} and \bref{for-3}, we have \bref{for-decom}. This ends the proof.
\qed

\begin{defi}
A {\bf multiplicity preserving triangular decomposition} of a polynomial system $\Sigma$ is a group of triangular sets $\{T_i^+,T_j^-,i=1,\ldots,m^+, j=1,\ldots,m^-\}$ in multiplicative variety sense such that
\begin{eqnarray}
\M(\Sigma)&=&\sum_{i=1}^{m^+}\M(T_i^+)-\sum_{j=1}^{m^-}\M(T_i^-).\label{eq-d}
\end{eqnarray}
\end{defi}
\noindent{\bf Remark:} We will show that a multiplicity preserving triangular sets decomposition exists for systems with two polynomials in the rest of the paper. Note that for a zero-dimensional polynomial system, the existence of \bref{eq-d} is obvious. The existence of \bref{eq-d} for for general case (dimension mixed, more polynomials) is our future work.

The following is a key result of the paper.
\begin{theorem}\label{thm-2}
Let $f_1,f_2\in K[x_1,\ldots,x_n,x]$ such that $\gcd$ $(f_1,f_2)=1$
and $\Cont(f_i,x)=1, i=1,2$. Then
\begin{equation}\label{eq-10}
\M(f_1,f_2)=\M(f_{k+1},f_{k+2})+\sum_{i=1}^k\M(g_i,f_{i+1})-\sum_{i=1}^k\M(m_{i},f_{i+1}),
\end{equation}
where $f_i,g_i,m_i$ are defined in Corollary \ref{cor-8}.
\end{theorem}
{\bf Proof.} From \bref{eq-9},  we have
$$\M(m_if_i,f_{i+1})=\M(f_{i+1},g_i\,f_{i+2}).$$
Then by Corollary \ref{for-2}, for $1\le i\le k$, we have
\begin{eqnarray}
&\M(m_i,f_{i+1})+\M(f_i,f_{i+1})=\M(g_i,f_{i+1})+\M(f_{i+1},f_{i+2}),& \label{eq-mul}\\
&\M(f_i,f_{i+1})=\M(f_{i+1},f_{i+2})+\M(g_i,f_{i+1})-\M(m_i,f_{i+1}).& \label{eq-mul}
\end{eqnarray}
So we have
\begin{eqnarray*}
\M(f_1,f_2)&=& \M(f_2,f_3)+\M(g_1, f_2)-\M(m_1,f_2)\\
           &=& \M(f_3,f_4)+\M(g_1, f_2)+\M(g_2, f_3)-\M(m_1,f_2)-\M(m_2,f_3)\\
           &&...\\
           &=& \M(f_{k+1},f_{k+2})+\sum_{i=1}^k\M(g_i,f_{i+1})-\sum_{i=1}^k\M(m_{i},f_{i+1}).\qed
\end{eqnarray*}
\noindent{\bf Remark:} The decomposition is about the
$(n-2)$-dimensional component of $\M(f_1,f_2)$.

 \begin{corollary} \label{cor-3}
Use the notations as Corollary \ref{cor-4}, we have
\begin{eqnarray}\label{eq-199}
\M(f_1,f_2)
&=&\M(f_{k+1},f_{k+2})+\sum_{i=1}^k(\M(p_i,f_{i+1})-\M(w_i,f_{i+1}))\nonumber\\
&&-\sum_{i=1}^{k-1}(\M(m_i,\frac{m_{i-1}}{w_i})+\M(m_i,p_i)-\M(m_i,q_i))-\M(m_k,f_{k+1}).
\end{eqnarray}
\end{corollary}
{\bf Proof.}
From Corollary \ref{cor-4}, we have $w_i\,g_i=m_{i-1}p_i$. So by \bref{for-1} and Corollary \ref{for-2}, we have
{\small
\begin{equation} \label{eq-g}
\M(g_i,f_{i+1})=\M(p_i,f_{i+1})+\M(m_{i-1},f_{i+1})-\M(w_i,f_{i+1}).
\end{equation}
}
This helps us simplifying the computation.
By \bref{eq-mul}, we have
{\small
\begin{eqnarray}\label{eq-general}
\M(f_1,f_2)&=&\M(f_{k+1},f_{k+2})+\sum_{i=1}^k(\M(p_i,f_{i+1})+\M(m_{i-1},f_{i+1})-\M(w_i,f_{i+1}))-\sum_{i=1}^k\M(m_{i},f_{i+1})\nonumber\\
&=&\M(f_{k+1},f_{k+2})+\sum_{i=1}^k\M(p_i,f_{i+1})-\sum_{i=1}^k\M(w_i,f_{i+1})
+\M(m_0,f_2)+\sum_{i=1}^{k-1}\M(m_{i},f_{i+2})\nonumber\\
&&-\sum_{i=1}^{k-1}\M(m_{i},f_{i+1})-\M(m_k,f_{k+1}).
\end{eqnarray}
}
From
$$m_i f_i+q_i f_{i+1} = \frac{m_{i-1}}{w_i} p_i f_{i+2},$$
we have
$$\M(m_i,q_i f_{i+1}) = \M(m_i,\frac{m_{i-1}}{w_i} p_i f_{i+2}).$$
By \bref{for-1} and Corollary \ref{for-2}, we have
\begin{eqnarray*}
\M(m_i, f_{i+1}) &=& \M(m_i, f_{i+2}) + \M(m_i,\frac{m_{i-1}}{w_i})+\M(m_i,p_i)-\M(m_i,q_i).
\end{eqnarray*}
And $m_0=1$, so $\M(m_0,f_2)=\emptyset$. Then we have \bref{eq-199}.
\qed

\noindent{\bf Remark:} The components $\M(m_i,\frac{m_{i-1}}{w_i})$,
$\M(m_i, p_i))$, and $\M(m_i, q_i)$ only involve polynomials in
$K[x_1,\ldots,x_n]$. Note that by Lemma \ref{lm-eud}, the
coefficient of $q_i$ in $x^t$ for $t>0$ is zero when $m_i=0$. These
components can also be decomposed into triangular sets recursively.
This corollary is very important since it provides a method to
eliminate the main variable $x$ in $f_i$'s, which simplifies the
decomposition. We can obtain another interesting phenomenon from simple observation, that is, the degree of all the resulting polynomials is bounded by the square of the degree of $f_1,f_2$.
\begin{example}
Consider the system $[f_1,f_2]=[{x}^{2}+{y}^{2}+{z}^{3}-1,x{z}^{2}-zy+1]$ with Corollary \ref{cor-3} under the variable order $x\prec y \prec z$,
\begin{eqnarray*}
f_3&=&{x}^{4}+{x}^{2}{y}^{2}-{x}^{2
}-xz-y+{y}^{2}z,\\
f_4&=&1-3\,{x}^{3}y-3\,x{y}^{3}+3\,xy+{x}^{2}
{y}^{3}+{y}^{5}-{y}^{3}+{x}^{7}+2\,{x}^{5}{y}^{2}-2\,{x}^{5}+{x}^{3}{y
}^{4}-2\,{x}^{3}{y}^{2}+{x}^{3}.
\end{eqnarray*}
$m_1={x}^{2},q_1=xz+y, m_0=w_1=p_1=1, m_2=( -x+{y}^{
2} )^{2},q_2=-{x}^{2}z+x{y}^{2}z+2\,xy-{y}^{3}-{x}^{5}-{x}^{3}{y}^{
2}+{x}^{3}.$ $w_2=p_2=1$. By Corollary \ref{cor-3}, we have the following decomposition.
\begin{equation*}
\M(f_1,f_2)=\M(f_4,f_3)+\M(m_1,q_1)-\M(m_2,f_3),
\end{equation*}
where $\M(m_1,q_1)=2\M(x,y)$ and
$\M(m_2,f_3)=2\M(x,y)+2\M(x-1,y-1)+2\M(h_1,h_2),$
where $h_1={x}^{6}+3\,{x}^{5}+2\,{x}^{4}+{x}^{2}+x+1$, $h_2=y-{x}^{4}-{x}^{3}+x^2$. Thus we have
$$\M(f_1,f_2)=\M(f_4,f_3)-2\M(x-1,y-1)-2\M(h_1,h_2).$$
Note that the component with negative multiplicity cannot be removed if using triangular form.
\end{example}

\section{Multiplicity Preserving Decomposition for System of Two Bivariate Polynomials}

In this section, we will consider the triangular decomposition of a
zero-dimensional bivariate polynomial system with two polynomials,
that is, $\Sigma=\{f,g\}\subset K[x,y]$. The method provided here is
complete for a zero dimensional bivariate polynomial system with two
polynomials. When $\Sigma$ is zero dimensional, $\M(\Sigma)$ defines
a multiplicative-zero set.

\subsection{Algorithm}
\begin{lemma}\label{cor-33}
Using the similar notations as Corollary \ref{cor-4}, if $\{f_1,f_2\}$ is zero-dimensional, we have
\begin{eqnarray}\label{eq-21}
\M(f_1,f_2)&=&\M(f_{k+1},f_{k+2})+\sum_{i=1}^k\M(p_i,f_{i+1})-\sum_{i=1}^k\M(w_i,f_{i+1})-\M(m_k,f_{k+1}).
\end{eqnarray}
\end{lemma}
{\bf Proof.} The lemma is a consequence of Corollary \ref{cor-3}.
Note that
$\M(m_i,\frac{m_{i-1}}{w_i})=\M(m_i,p_i)=\M(m_i,q_i)=\emptyset$
since $\gcd(m_i,$ $p_i)$, $\gcd(m_i,\frac{m_{i-1}}{w_i})$,
$\gcd(m_i,q_i)$ are constant.

The following corollary is useful.
\begin{corollary} \label{cor-positive}
If  $w_i(1\le i \le k)$ are constants and
$f_{k+1}=l_1(x)\,y^t+l_0(x)$ for $t>0$ and $l_0,l_1\in K[x]$ in
\bref{eq-21}, we have
\begin{eqnarray}\label{eq-11}
\M(f_1,f_2)=\M(f_{k+1},f_{k+2})+\sum_{i=1}^k\M(p_i,f_{i+1}).
\end{eqnarray}
Furthermore, if $p_i (1\le i\le k)$ are constant, we have
\begin{eqnarray}\label{eq-12}
\M(f_1,f_2)=\M(f_{k+1},f_{k+2}).
\end{eqnarray}
\end{corollary}
{\bf Proof.}

Since $f_{k+1}=l_1(x)\,y^t+l_0(x)$ and $\Cont(f_{k+1},y)=1$,
$\M(m_k,f_{k+1})=\emptyset$. Note that $m_k$ is a factor of
$l_{1}^{n}$ for some positive integer $n$. Thus
$\M(m_0,f_2)=\emptyset$, $\M(m_k,f_{k+1})=\emptyset$. So from
\bref{eq-21}, we have \bref{eq-11}. If $p_i=1$, \bref{eq-12} is a
consequence of \bref{eq-11}. \qed

Now we will consider the complexity of our method under the condition of
Corollary \ref{cor-positive}. We consider this case because it is
usually the case for almost all zero-dimensional bivariate
polynomial system with two polynomials. So the result is
interesting. At first, we need to introduce some notations, which
can be found in \cite{det}. Let $\mathcal{L}(f)$ bound the bitsize
of the coefficients of $f\in K[x,y]$ (including a bit for the sign).
We assume {\rm lg}$(\deg(f))=\mathcal{O}(\mathcal{L}(f))$.
 For $a\in \Q$, $\mathcal{L}(a)$ is the maximum bitsize of $a$'s numerator and denominator. Let $M(\tau)$ denote
the bit complexity of multiplying two integers of size $\tau$, and
$M(d,\tau)$ the complexity of multiplying two univariate polynomials
of degrees $\le d$ and coefficient bitsize $\le \tau$. Using FFT,
$M(\tau)=\widetilde{\mathcal{O}}_B(\tau)$  and
$M(d,\tau)=\widetilde{\mathcal{O}}_B(d\tau)$.

\begin{lemma}\cite{det,re} \label{lem-gcd}
Let $f,g\in \Z[x]$, $\deg(f), \deg(g)\le n$, and $\mathcal{L}(f), \mathcal{L}(g)\le \tau$. We can compute $\gcd(f,g)$ in $\widetilde{\mathcal{O}}_B(n^2\tau)$.
\end{lemma}

\begin{lemma}\cite{det,re} \label{lem-srq}
Let $f,g\in \Z[x,y]$, $\deg(f), \deg(g)\le n$, and $\mathcal{L}(f), \mathcal{L}(g)\le \tau$. We can compute the subresultant sequence of $f$ and $g$ in $\widetilde{\mathcal{O}}_B(n^6\tau)$.
\end{lemma}

\begin{theorem}
If  $w_i(1\le i \le k)$ are constant and $f_{k+1}=l_1(x)\,y^t+l_0(x)$ for some positive integer $t$, $l_0,l_1\in K[x]$ in \bref{eq-21}, and $K=\Z$, we can decompose a zero-dimensional bivariate system with two polynomials into multiplicity preserving triangular sets in $\widetilde{\mathcal{O}}_B(n^7\tau)$.
\end{theorem}
{\bf Proof.} We can compute a subresultant sequence of $f$ and $g$
at first. It can be computed in $\widetilde{\mathcal{O}}_B(n^6\tau)$
by Lemma \ref{lem-srq}. Then we simplify each pseudo-division step
to derive (\ref{eq-18}) from the highest degree of the sequence in
$y$ to the lowest degree. Let $\{F_1,\ldots,F_{k+2}\}$ be the
subresultant sequence of $f$ and $g$. We need only consider the case
of regular subresultant sequence since the complexity of the regular
case also bounds the degenerate case. Consider the formula
\begin{equation}\label{eq-psd}
l_{i+1}^2 F_i+Q_i F_{i+1}=l_i^2F_{i+2}.
\end{equation}
Assume that we have computed the contents of $F_i$ and $F_{i+1}$,
say $r_i, r_{i+1}$. $F_1,F_2$ are $f_1, f_2$. And the contents of
$f$ or $g$ can be computed in $\widetilde{\mathcal{O}}_B(n^3\tau)$
by Lemma \ref{lem-gcd}, which can be ignored comparing to
$\widetilde{\mathcal{O}}_B(n^6\tau)$. For each $F_i$, it is well
known that $\deg(F_i)\le n^2, \mathcal{L}(F_i)=\mathcal{O}(n\tau)$
(for reference see \cite{det}). And $\deg(F_i,y)\le n-1$ for $i\ge
3$. Thus for any coefficient of $F_i$, say $h\in \Z[x]$, we have
$\deg(h)\le n^2, \mathcal{L}(h)=\mathcal{O}(n\tau)$. So to compute
the content of $F_i$ with $\deg(F_i,y)=h$, we need to compute at
most $h$ $\gcd$ each in $\widetilde{\mathcal{O}}_B(n^5\tau)$. Let
$r_{i+2}=\Cont(F_{i+2},y)$. In order to derive \bref{eq-18}, we need
to delete $\gcd(l_{i+1}^2 F_i,Q_i F_{i+1},l_i^2F_{i+2})$ from the
two side of \bref{eq-psd}. So we need to bound $\gcd(l_{i+1}^2\,
r_i,l_i^2\,r_{i+2})$.  Note that $\deg(s)\le n^2,
\mathcal{L}(s)=\mathcal{O}(n\tau)$ holds for $s=l_i$ or $s=l_{i+1}$
and we can not optimize the degree of $l_{k+1}$. But we can compute
$r=\gcd(l_i,l_{i+1})$, which is bounded by
$\widetilde{\mathcal{O}}_B(n^5\tau)$. And
$\gcd((\frac{l_i}{r})^2,r_{i+2})$ can be bounded by
$2\,\widetilde{\mathcal{O}}_B(n^5\tau)$ as below. We can compute
$w=\gcd((\frac{l_i}{r}),r_{i+2})$, and then
$\gcd((\frac{l_i}{r}),\frac{r_{i+2}}{w})$.
$\gcd((\frac{l_{i+1}}{r})^2,r_{i})$ is also bounded by
$2\,\widetilde{\mathcal{O}}_B(n^5\tau)$. So to obtain \bref{eq-18}
from \bref{eq-psd} for $\deg(F_i,y)=k$, we need
$(k+5)\widetilde{\mathcal{O}}_B(n^5\tau)$. Then we can decide $m_i,
g_i$ in \bref{eq-8} by two divisions. Since $w_i$'s are constant,
$m_{i-1}|g_i$. Thus, we can obtain $p_i$ by one division. When $h$
changes from $n$ to $1$, we can bound it by
$\frac{n^2+9n}{2}\widetilde{\mathcal{O}}_B(n^5\tau)$, that is,
$\widetilde{\mathcal{O}}_B(n^7\tau)$. Then the total complexity is
$\widetilde{\mathcal{O}}_B(n^7\tau)$. \qed

\noindent{\bf Remark:} For many $f, g\in K[x,y]$, the last two elements of the subresultant sequence $F_{k+1},F_{k+2}$ form a multiplicity preserving triangular decomposition of $f,g$. Thus, we can compute the decomposition in $\widetilde{\mathcal{O}}_B(n^6\tau)$.

The lemma gives a multiplicity preserving
triangular decomposition of a bivariate polynomial system. But there
exist some triangular sets with negative multiplicities. The following
results gives an algorithm to remove the triangular sets with
negative multiplicities.

\begin{theorem}\label{cor-main}
There exists an algorithm to decompose a zero-dimensional bivariate polynomial system $\{f_1,f_2\}\subset K[x,y]$ into a set of triangular sets, such that
\begin{equation} \label{eq-decom}
\M(f_1,f_2)=\sum_{i=1}^N\M(g_i,h_i),
\end{equation}
where $g_i\in K[x], h_i\in K[x,y]$.
\end{theorem}
{\bf Proof.}
In the case $f_1=h_1 f_1', f_2=h_2 f_2'$ having factors in $K[x]$ but $\gcd(f_1,f_2)=1$, where $h_i=\Cont(f_i,y), i=1,2$, we have
\begin{eqnarray}\label{eq-14}
\M(f_1, f_2)
=\M(h_1,f_2')+\M(f_1',h_2)+\M(f_1',f_2').
\end{eqnarray}
Since $\gcd(h_1\,f_1',h_2\,f_2')=1$, $\M(h_1,h_2)=\emptyset$.

By \bref{eq-14}, we can assume that $\Cont(f_i,y)=1, i=1,2$.
From Lemma \ref{cor-33}, we have \bref{eq-21}. We put the triangular sets on the righthand side of \bref{eq-21} into two sets: $W_1=\M(f_{k+1},f_{k+2})+\sum_{i=1}^k\M(p_i,f_{i+1})$, $W_2=\sum_{i=1}^k\M(w_i,f_{i+1})+\M(m_k,f_{k+1})$. It is clear that $W_2\subset W_1$. Our aim is to delete the multiplicative-zeros of $W_2$ from $W_1$ to derive a group of triangular sets with positive coefficients.

Take any triangular set $U=(u_1(x),u_2(x,y))$ out of $W_2$, we can
compute gcd of $u_1(x)$ and some $v_1(x)$, where
$V=(v_1(x),v_2(x,y))$ is one triangular set of $W_1$. Denote the gcd
as $p(x)$. Then decomposing $U$, we have $(p(x),u_2(x,y))$,
and put $(\frac{u_1(x)}{p(x)},u_2(x,y))$ into $W_2$ again. Decomposing
$V$, we have $(p(x),v_2(x,y))$. And put $(\frac{v_1(x)}{p(x)},v_2(x,y))$
into $W_1$. We can compute the $\gcd$, say $w(x,y)$, of $u_2(x,y)$ and
$v_2(x,y)$ modulo $p(x)$, then remove the factor $w(x,y)$ from
$u_2(x,y)$ ($v_2(x,y)$) and put the left part into $U$($V$). We can
also use the method in \cite{sun} to decompose $(p(x),u_2(x,y))$ and
$(p(x),v_2(x,y))$ into irreducible and regular triangular sets, and
then we can easily decide that whether two irreducible and regular
triangular sets have same zero set or not. Thus we can remove the
triangular sets in $W_2$. In the end, we can remove all the zero
sets in $W_2$. We prove the theorem.\qed

We will give a multiplicity preserving algorithm to decompose a bivariate polynomial system into triangular sets based on the theory above.
\begin{alg} \label{alg-main}Input: a zero-dimensional bivariate polynomial system $\P_1=\{f_1(x,y),f_2(x,y)\}\in K[x,y]$, and $d_1=\deg(f_1,y)\ge d_2=\deg(f_2,y)$.
Output:  a group of triangular sets
$\P=\{[g_i(x),h_i(x,y)],i=1,...,n\}$ such that
$\M(f_1,f_2)=\Sigma_{i=1}^n\M(f_i,h_i)$.
\end{alg}
\begin{enumerate}
\item $\M_p=\emptyset, \M_n=\emptyset$.

\item 
Compute $h_i=\Cont(f_i,y), i=1,2$. Let $f_i=f_i/h_i$.
$\M_p=\M_p\cup\{[h_1,f_2],[h_2,f_1]\}.$

\item Let $m_0=1$. While $\deg(f_2,y)>0$, do
         \begin{itemize}
         \item By pseudo-division, we have $m_1\,f_1+q_1 f_2=f_3$ and $h=\Cont(f_3,x)$.
         $f_3=\frac{f_3}{h}, v=\gcd(m_1,h)$, $m_1=\frac{m_1}{v}, h=\frac{h}{v}$.
         Let $q=\gcd(m_0,h)$, $w=\frac{m_0}{q}$, $p=\frac{h}{q}$. If $p$ is not a constant, $\M_p=\M_p\cup\{[p,f_2]\}$. If $w$ is not a constant, $\M_n=\M_n\cup\{[w,f_2]\}.$
         \item $f_1=f_2,f_2=f_3,m_0=m_1.$
         \end{itemize}
        If $\deg(f_1,y)>1$, $\M_n=\M_n\cup\{[m_0,f_1]\}.$
\item
If $\M_n$ is not empty, following the method in the proof of Theorem
\ref{cor-main}, we can remove the multiplicative varieties in
$\M_n$ from $\M_p$. Thus we obtain a group of triangular sets as
\bref{eq-decom}.
\end{enumerate}
{\bf Proof.} The termination of the algorithm is clear since the degree of $f_1$ and $f_2$ is finite. The correctness of the algorithm is guaranteed by Theorem \ref{cor-main} and Lemma \ref{cor-33}.

\begin{example}
Let $\C$ be the curve defined by
$$f=2\,{y}^{4}-3\,{y}^{2}x+{x}^{2}-2\,{x}^{3}+{x}^{4}.$$
We will compute the $y$-critical points ($f=\frac{\partial f}{\partial y}=0$) of $\C$.
$$f_y=\frac{\partial f}{\partial y}=8\,{y}^{3}-6\,yx.$$
Delete the content $2$, $f_y=\frac{f_y}{2}$. In the following, we will solve the system $\Sigma=\{f,f_y\}$. Following our algorithm, we have
$$m_1\,f+q_1\,f_y=p_1\,f_3,$$
where $m_1=4,p_1=x,q_1=-y,f_3=-3\,{y}^{2}+2\,x-4\,{x}^{2}+2\,{x}^{3}$.
$$m_2\,f_y+q_2\,f_3=p_2\,f_4,$$
where $m_2=3,q_2=4\,y,p_2=x\, (-1-16\,x+8\,{x}^{2} ),f_4=y$. Note that here $w_2=4$. We ignore it since it is a constant.
$$m_3\,f_3+q_3\,f_4=f_5,$$
where $m_3=1,q_3=3\,y,f_5=2\,x\,(x-1)^2$. Similarly, we ignore $w_3=3$. With Corollary \ref{cor-positive}, we have
$$\M(f,f_y)=\M(f_5,f_4)+\M(p_1,f_y)+\M(p_2,f_3).$$
We can find that
\begin{eqnarray*}
\M(f_5,f_4)&=&\M(x,y)+2\M(x-1,y),\\
\M(p_1,f_y)&=&\M(x,4\,{y}^{3}-3\,yx)=3\M(x,y),\\
\M(p_2,f_3)
&=&2\M(x,y)+\M(-1-16\,x+8\,x^2, -4\,y^2+3\,x).
\end{eqnarray*}
We find that $\M(x,y)$ and $\M(x-1,y)$ are zeros with multiplicities 6, 2, respectively. And the other zeros are with multiplicities 1.
\end{example}

\subsection{Implementation and Comparison}

We implement our algorithm in Maple. We compare the computing time
of our methods with several other related methods. One is the
regular chains method \cite{dahan,mmm} (package RegularChains in Maple
13, including two functions), one is Characteristic set method in Epsilon\cite{wang1}, the other is a package  wsolve (see \cite{wsolve}). All the results
are collected on a PC with a 3.2GHz CPU, 2.00G memory, and running
Microsoft Windows XP. We use Maple 13 in the experiments.

We run 100 examples in each case and compute their average
computing time in Table 1. We take random dense polynomials with coefficients bounded by $[-100,100]$ for each example. Table 1 is the timings for the given methods in seconds. Here MPTD means the method provided in this paper,  RC (BMT) means regular
chains method of function ``Triangularize''(``BivariateModularTriangularize"), CS means characteristic set method (``charsets") in Epsilon, and WS means wsolve method. The first row
is $[\deg(f_1),\deg(f_2)]$. The first column represents the methods. ``-" means out of memory or we did not test it.

\begin{table}
{\tiny
\begin{center} \begin{tabular}{|c| c | c | c | c | c |c|}
\hline
degree& [5,4]  & [7, 5] &  [9, 7] &  [13, 11] & [23, 21]& [33,31]\\ \hline
MPTD  &0.006 & 0.019 & 0.105& 1.363& 62.894&884.577\\\hline

BMT &0.014 & 0.024 & 0.036& 0.077& 0.280&0.848\\\hline

RC &0.194 & 0.343 & 0.810& 4.520& 127.194&1075.653\\\hline

CS &0.082 & 0.880 & 25.101& -& - &-\\\hline

WS &0.134 & 3.881& 410.399& - & - &-\\ \hline

\end{tabular}
\end{center}
\caption{Timings for different methods}
}
\end{table}

We need to mention that only MPTD can compute the multiplicities of the zeros of the bivariate polynomial system. BMT and RC are implemented by C code but other's are in Maple. MPTD and BMT are only for bivariate polynomial system but other methods work for general system. Note that with a mirror modification, MPTD can multiplicity preserving triangular decompose system with two multivariate polynomials.

We can conclude from the table that MPTD and BMT are always faster than RC, CS and WS. For the system with low degrees, CS, WS are a little faster than RC, but they both are very slow when the degree of the system more than 10. MPTD is always a little faster than RC. MPTD is a little faster than BMT for low degree system, but much slower than BMT for systems with high degrees. There are several reasons: BMT is in C, using modular method and using FFT based arithmetic, but MPTD does not use these techniques.

\section{Conclusion}
We present an algorithm to decompose a polynomial system with two
polynomials into triangular sets. Different from the existing
methods for triangular decomposition, our method preserves
the multiplicity of the zeros  or components of the systems. We
implement the method for bivariate polynomial systems. We will extend
the method to the systems with more polynomials in the future.

\section{Acknowledgement} The work is partially supported by National Key Basic Research Project of China and China-France cooperation project EXACTA. The first author is partially supported by NSFC Grant 11001258.

{}

\end{document}